\documentclass{appolb}
\usepackage[dvips]{graphicx}
\begin{document}
\title  {{\bf Nonextensive critical effects in the Nambu--Jona-Lasinio model
}\footnote{Presented (by JR) at the XXXI Mazurian Lakes Conference
on Physics, Piaski, Poland, August 30--September 5, 2009}
\author {Jacek Ro\.zynek and Grzegorz Wilk}
\address {Theoretical Physics Division, So{\l}tan Institute
for Nuclear Studies,\\
Ho{\.z}a 69, 00-681 Warsaw, Poland}}

\maketitle

\begin{abstract}

The critical phenomena in strongly interaction matter are
generally investigated using the mean-field model and are
characterized by well defined critical exponents. However, such
models provide only average properties of the corresponding order
parameters and neglect altogether their possible fluctuations.
Also the possible long range effect are neglected in the mean
field approach. Here we investigate the critical behavior in the
nonextensive version of the Nambu Jona-Lasinio model (NJL). It
allows to account for such effects in a phenomenological way by
means of a single parameter $q$, the nonextensivity parameter. In
particular, we show how the nonextensive statistics influence the
region of the critical temperature and chemical potential in the
NJL mean field approach.

\end{abstract}


\section{\label{sec:I}Introduction}

Critical phenomena in strongly interaction matter are of great
interest nowadays, cf., for example, \cite{HK,SFR}. They are
usually described by a mean field type of theories\footnote{See,
for example, \cite{SW} or \cite{CRS} mentioned in this
presentation, we refer to them for the most recent literature on
this subject.}. Such theories are based on the usual
Boltzmann-Gibbs (BG) statistical mechanics and reflect only
behavior of the mean field, i.e., are not able to accommodate
effects of possible fluctuations and/or correlations caused, among
others, by the smallness of the sample of matter under
consideration, by its rapid evolution, or by limitations of the
available phase space. All these factors (both separately and
taken together) render the spatial configuration of the system
being far from uniform and prevent the global equilibrium from
being established. Nevertheless, it is know in the literature that
it is possible to maintain simplicity of the statistical
description and, at he same time, to account for these effects
provided one uses a nonextensive version of the statistical
mechanics, for example the one proposed by Tsallis \cite{T} (which
we shall use here). In this approach action of all the above
mentioned factors is summarily accounted for by one additional
parameter $q$, the nonextensivity parameter. It does not
differentiate between the particular dynamical phenomena
responsible for departure from the BG picture. In this approach
the usual BG exponent, $\exp(-X/T)$, is deformed into the so
called $q$-exponent (Tsallis distribution), $\exp_q(-X/T) =
[1-(1-q)X/T]^{1/(1-q)}$, such that for $q \rightarrow 1$ one
recovers the BG picture again.

The applications of the nonextensive statistical mechanics to
nuclear and particle physics are numerous and we refer to
\cite{EPJA} for details. In what follows we shall present the
nonextensive version of the NJL model, the $q$-NJL model \cite{RW}
\footnote{ Nonextensive calculations using the Walecka model for
dense nuclear matter has been done in \cite{PSA}.}. Details of the
$q$-NJL model and most of results were already presented in
\cite{RW}, here we shall concentrate on the influence of dynamical
factors causing nonextensivity and represented by parameter $q$ on
the vicinity of the {\it critical end point} (CEP).

\section{\label{sec:II}Results}

Let us present first the basic elements of the $q$-NJL model
introduced in \cite{RW} (to which we refer for more details). It
is a $q$-version of standard $SU(3)$ NJL model with $U(1)_A$
symmetry described in \cite{CRS}, with the usual Lagrangian of the
NJL model used in a form suitable for the bosonization procedure
(with four quarks interactions only), from which we obtain  the
gap equations for the constituent quark masses $M_i$:
\begin{eqnarray}
 M_i = m_i - 2g_{_S} \big <\bar{q_i}q_i \big > -2g_{_D}\big
 <\bar{q_j}q_j\big > \big <\bar{q_k}q_k \big >\,,\label{gap}
 \end{eqnarray}
with cyclic permutation of $i,j,k =u,d,s$ and with the quark
condensates given by $\big <\bar{q}_i q_i \big > = -i \mbox{Tr}[
S_i(p)]$ ($S_i(p)$ is the quark Green function); $m_i$ denotes the
current mass of quark of flavor $i$. We consider a system of
volume $V$, temperature $T$ and the $i^{th}$ quark chemical
potential $\mu_i$ characterized by the baryonic thermodynamic
potential of the grand canonical ensemble (with quark density
equal to $\rho_i = N_i/V$, the baryonic chemical potential $\mu_B=
\frac{1}{3} (\mu_u+\mu_d+\mu_s)$ and the baryonic matter density
as $\rho_B = \frac{1}{3}(\rho_u+\rho_d+\rho_s)$),
\begin{equation}
\Omega (T, V, \mu_i )= E- TS - \sum_{i=u,d,s} \mu _{i} N_{i} .
\label{tpot}
\end{equation}
The internal energy, $E$, the entropy, $S$, and the particle
number, $N_i$, are given by \cite{CRS} (here $E_i = \sqrt{M_i^2 +
p^2}$):
\begin{eqnarray}
E &=&- \frac{ N_c}{\pi^2} V\sum_{i=u,d,s}\left[
   \int p^2 dp  \frac{p^2 + m_{i} M_{i}}{E_{i}}
   (1 - n_{i}- \bar{n}_{i}) \right] - \nonumber\\
   && - g_{S} V \sum_{i=u,d,s}\, \left(\big <
\bar{q}_{i}q_{i}\big > \right)^{2}
   - 2 g_{D}V \big < \bar{u}u\big > \big < \bar{d}d\big > \big <
\bar{s}s\big > , \label{energy} \\
 S &=& \! -\frac{ N_c}{\pi^2} V \sum_{i=u,d,s}\int p^2 dp \cdot
 \tilde{S}, \label{entropy}\\
 && {\rm where}\quad \tilde{S} =  \bigl[ n_{i} \ln n_{i}+(1-n_{i})\ln (1-n_{i})
   \bigr]\!\! +\!\! \bigl[ n_{i}\rightarrow 1 - \bar n_{i} \bigr],\nonumber\\
N_i &=& \frac{ N_c}{\pi^2} V \int p^2 dp
  \left( n_{i}-\bar n_{i} \right) \label{number}.
\end{eqnarray}
The $n_{i}= 1/\left\{ \exp\left[\beta \left(E_{i} -
\mu_{i}\right)\right] + 1\right\}$ and $\bar n_{i} =
1/\left\{\exp\left[ \left( \beta(E_{i} + \mu_{i} \right)\right] +
1\right\}$ are, respectively, quark and antiquark occupation
numbers with which one calculates values of the quark condensates
present in Eq. (\ref{gap}),
\begin{eqnarray}
\big <\bar{q}_i  q_i \big> \!= \!\ - \frac{ N_c}{\pi^2} \!\!\!
\sum_{i=u,d,s}\left[ \int \frac{p^2M_i}{E_i} (1\,-\,n_{i}-\bar
n_{i})\right]dp .\label{gap1}
\end{eqnarray}
Eqs. (\ref{gap}) and (\ref{gap1}) form a self consistent set of
equations from which one gets the effective quark masses $M_i$ and
values of the corresponding quark condensates.

The values of the pressure, $P$, and the energy density,
$\epsilon$, are defined as:
\begin{equation} \label{p}
 \!\!\!  P(\mu_i, T)\! =\! - \frac{\Omega(\mu_i, T)}{V},~~
   \epsilon(\mu_i, T)\! =\! \frac{E(\mu_i, T)}{V}~~{\rm
   with}~~P(0,0)=\epsilon(0,0)=0.
\end{equation}
The $q$-statistics is introduced by using the $q$-form of quantum
distributions for fermions $(+1)$ and bosons $(-1)$, namely,
$n_{qi} = 1/\left\{\tilde{e}_q(\beta(E_i - \mu_i))\pm 1]\right\}$
where ($x = \beta(E -\mu)$) $\tilde{e}_q(x)=\left[ 1 +
|(q-1)x|\right]^{x/|(q-1)x|}$ \cite{RW}. With such choice one can
treat consistently on the same footing quarks and antiquarks,
which should show the particle-hole symmetry observed in the
$q$-Fermi distribution in plasma containing both particles and
antiparticles, namely that $n_q(E,\beta,\mu,q) = 1 - n_{2-q}(-
E,\beta,-\mu)$ \footnote{ In a system containing both particles
and antiparticles both $q$ and $2 - q$ occur (i.e., one can
encounter both $q > 1$ and $q <1$ at the same time). It means that
not only the $q > 1$ but also $ q < 1$ (or $(2 - q) > 1$ have
physical meaning in the systems we are considering what differs
our $q$-NJL model from the $q$-version of the QHD-I model
presented in \cite{PSA}. Notice that for $q\rightarrow 1$ one
recovers the standard FD distribution, $n(\mu,T)$. It is important
to realize that for $T\rightarrow0$ one always gets
$n_q(\mu,T)\rightarrow n(\mu,T)$, irrespectively of the value of
$q$ \cite{PSA}, i.e., we can expect any nonextensive signature
only for high enough temperatures.}. The $q$-NJL model is obtained
by replacing the formulas of Section \ref{sec:I}  with their
$q$-counterparts in what concerns the form of the FD
distributions.  Additionally, when calculating energies and
condensates we follow \cite{AL} and use the $q$-versions of
energies and quark condensates replacing Eqs. (\ref{energy}) and
(\ref{gap1}) by:
\begin{eqnarray}
E_q\! &=&\! - \frac{ N_c}{\pi^2} V\!\!\!\sum_{i=u,d,s}\!\!\left[
   \int p^2 dp  \frac{p^2 + m_{i} M_{i}}{E_{i}}
   (1 - n^q_{qi}- \bar{n}^q_{qi}) \right] - \nonumber\\
   &&\!\! - g_{S} V\!\!\!\sum_{i=u,d,s}\!\! \left(\big <
\bar{q}_{i}q_{i}\big >_q \right)^{2}
   - 2 g_{D}V \big < \bar{u}u\big >_q \big < \bar{d}d\big >_q \big <
\bar{s}s\big >_q , \label{q_energy}
\end{eqnarray}
and
\begin{equation}
 \big <\bar{q}_i
q_i \big>_q \! = \!\ - \frac{ N_c}{\pi^2} \!\!\!
\sum_{i=u,d,s}\left[ \int \frac{p^2M_i}{E_i} (1\,-\,n^q_{qi}-
\bar{n}^q_{qi})\right]dp .\label{q_gap1}
\end{equation}

\begin{figure}[t]
  \begin{center}
   \includegraphics[width=6.cm]{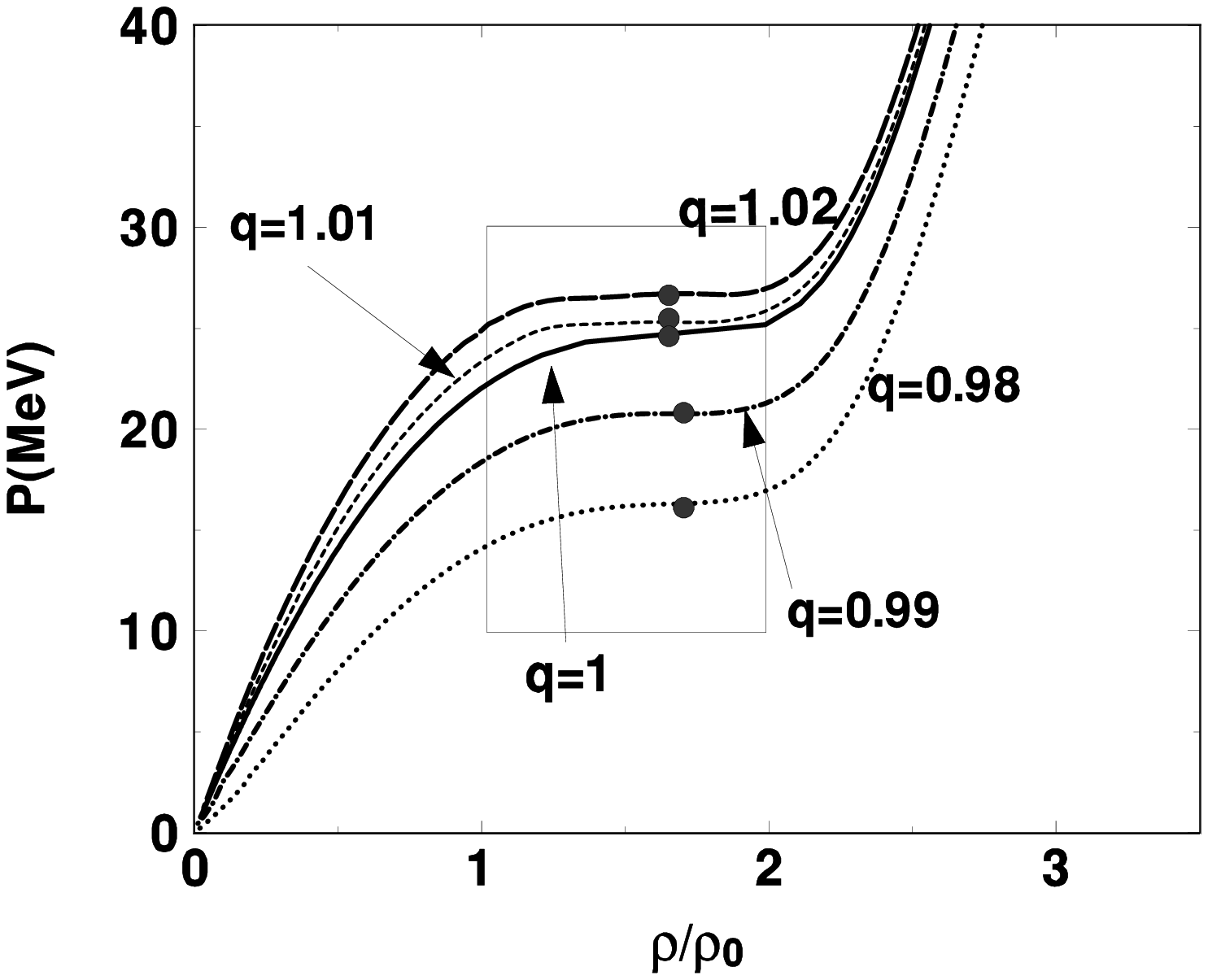}
   \includegraphics[width=6.cm]{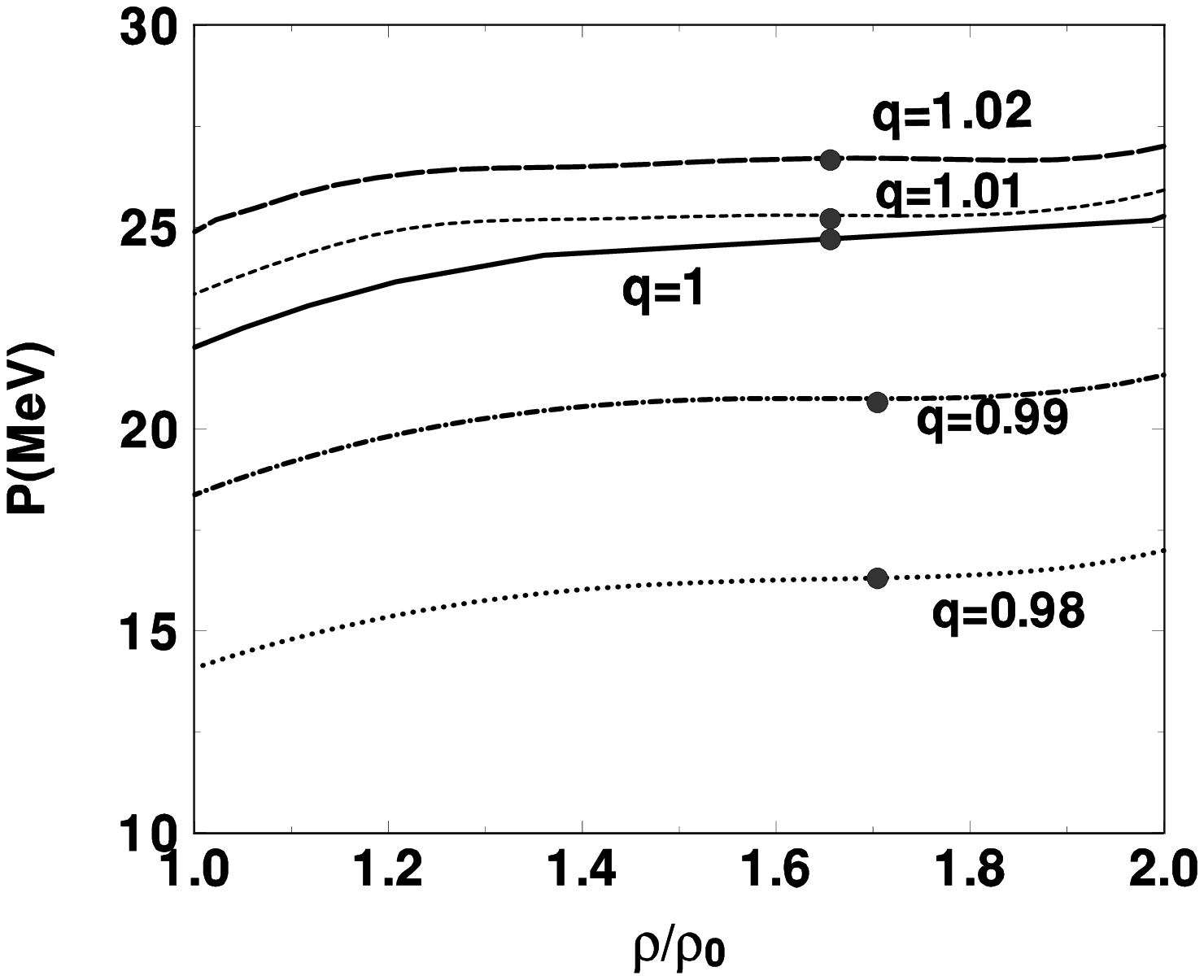}
   \caption{The pressure at critical temperature $T_{cr}$ as a
            function of compression $\rho/\rho_0$ calculated
            for different values of the nonextensivity parameter
            $q$ (the area marked at the left panel is shown in
            detail at the right panel).
            The dots indicate positions of the inflection
            points for which first derivative of pressure in
            compression vanishes. As in \cite{CRS} for $q = 1$
            the corresponding compression is $\rho/\rho_0 = 1.67$
            (and this leads to $\mu = 318.5$ MeV); it remains the
            same for $q > 1$ considered here (but now $\mu = 321$
            MeV for $q = 1.01$ and $\mu = 326.1$ MeV for $q = 1.02$)
            whereas it is shifted to $\rho/\rho = 1.72$ for
            $ q< 1$ ($\mu = 313$ MeV for $q = 0.99$
            and $\mu = 307.7$ MeV for $q = 0.98$).}
   \label{Figure1}
  \end{center}
\end{figure}

On the other hand, again following \cite{AL}, densities which are
given by the the $q$-version of Eq. (\ref{number}) are calculated
with $n_q$'s (not with $n_q^q$, as in (\ref{q_energy}) and in
(\ref{q_gap1})). The pressure for given $q$ is calculated using
the above $E_q$ and the $q$-entropy version of Eq. (\ref{entropy})
with (cf. \cite{TPM})
\begin{equation}
\tilde{S}_q\!\! =\!\! \left[ n^q_{qi} \ln_q n_{qi} +
(1-n_{qi})^q\ln_q (1-n_{qi}) \right] + \left\{ n_{qi}\rightarrow
1\! -\! \bar n_{qi} \right\}. \label{q entropy}
\end{equation}

\begin{figure}[t]
  \begin{center}
   \includegraphics[width=5.6cm]{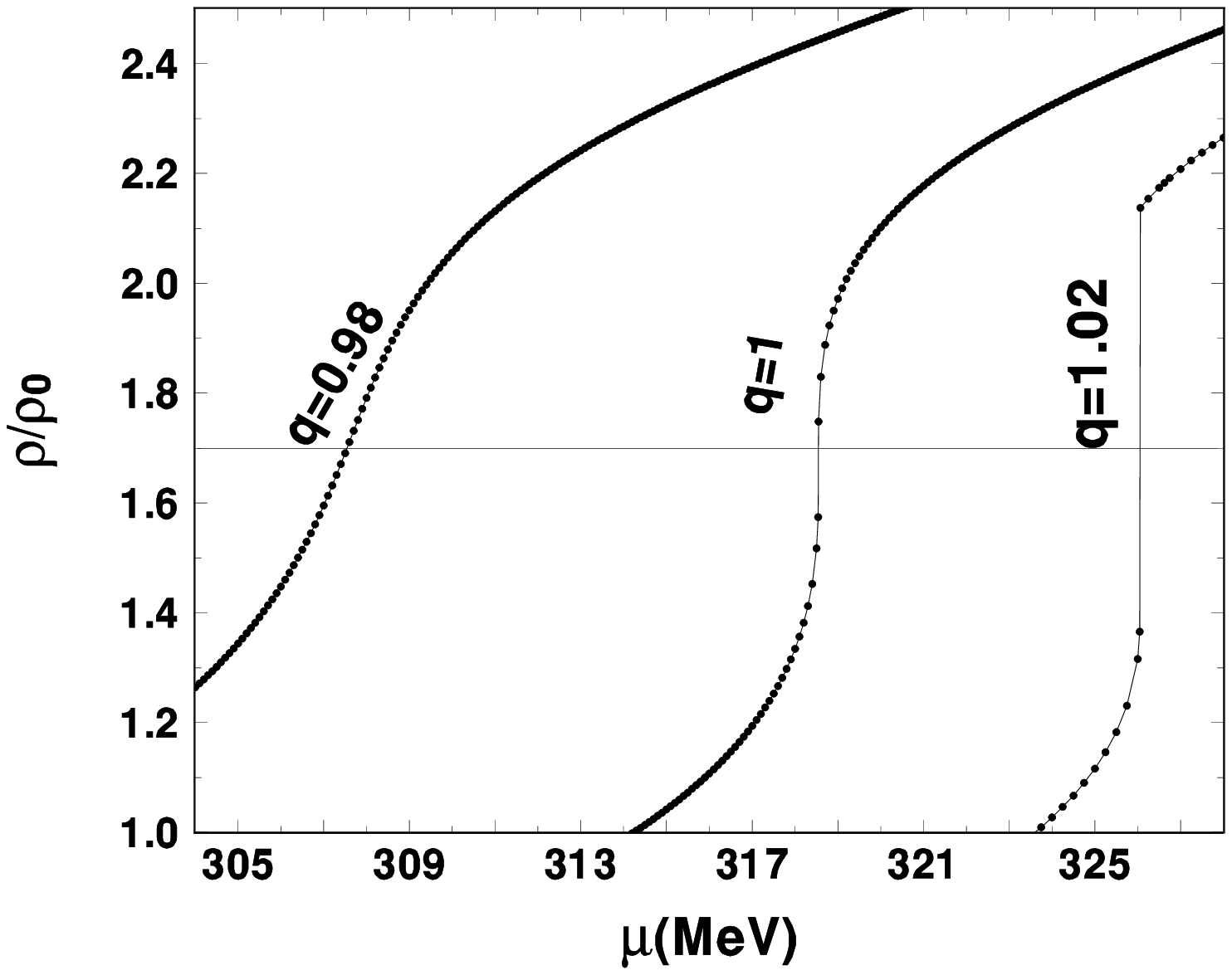}
   \includegraphics[width=5.6cm]{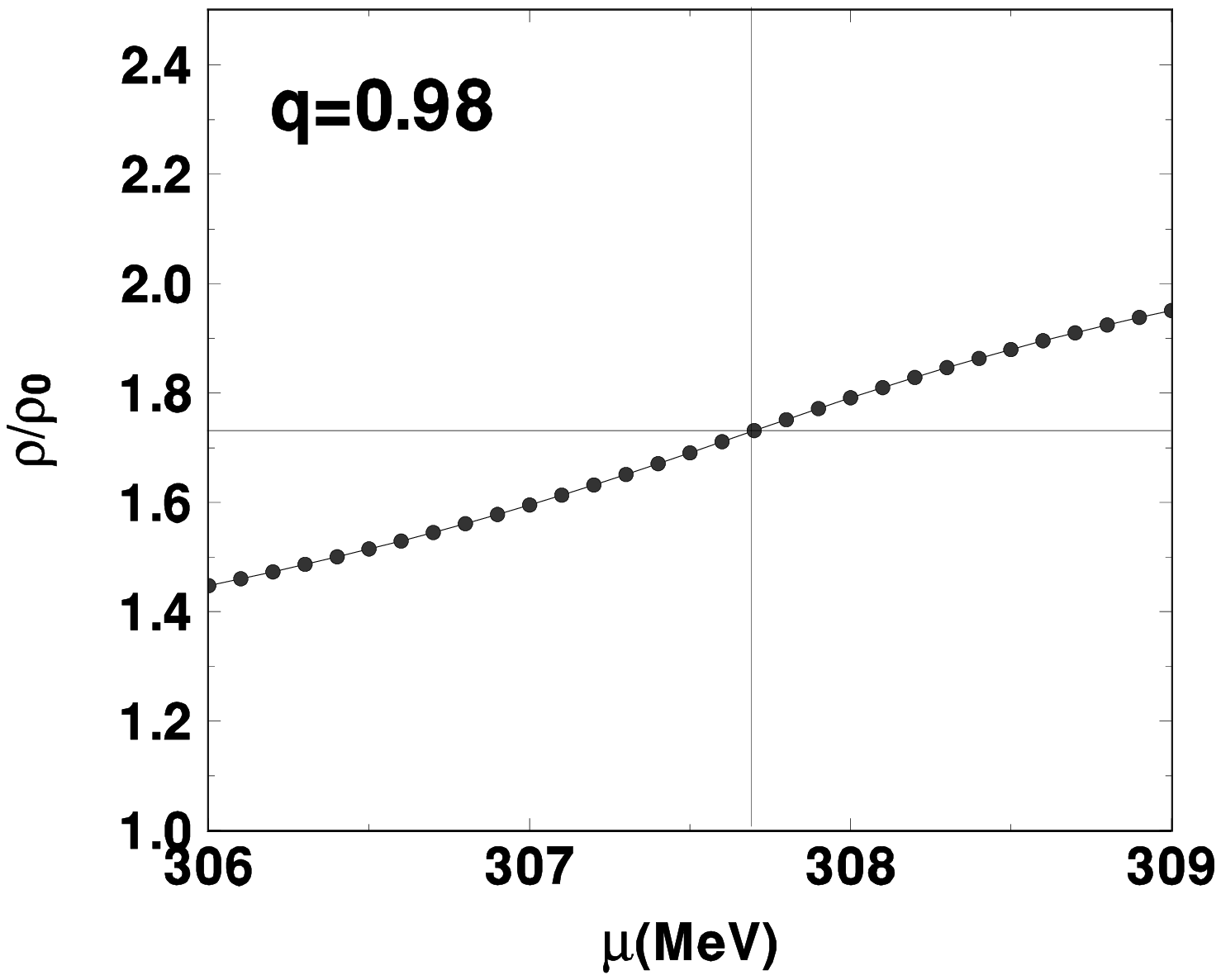}
   \includegraphics[width=5.6cm]{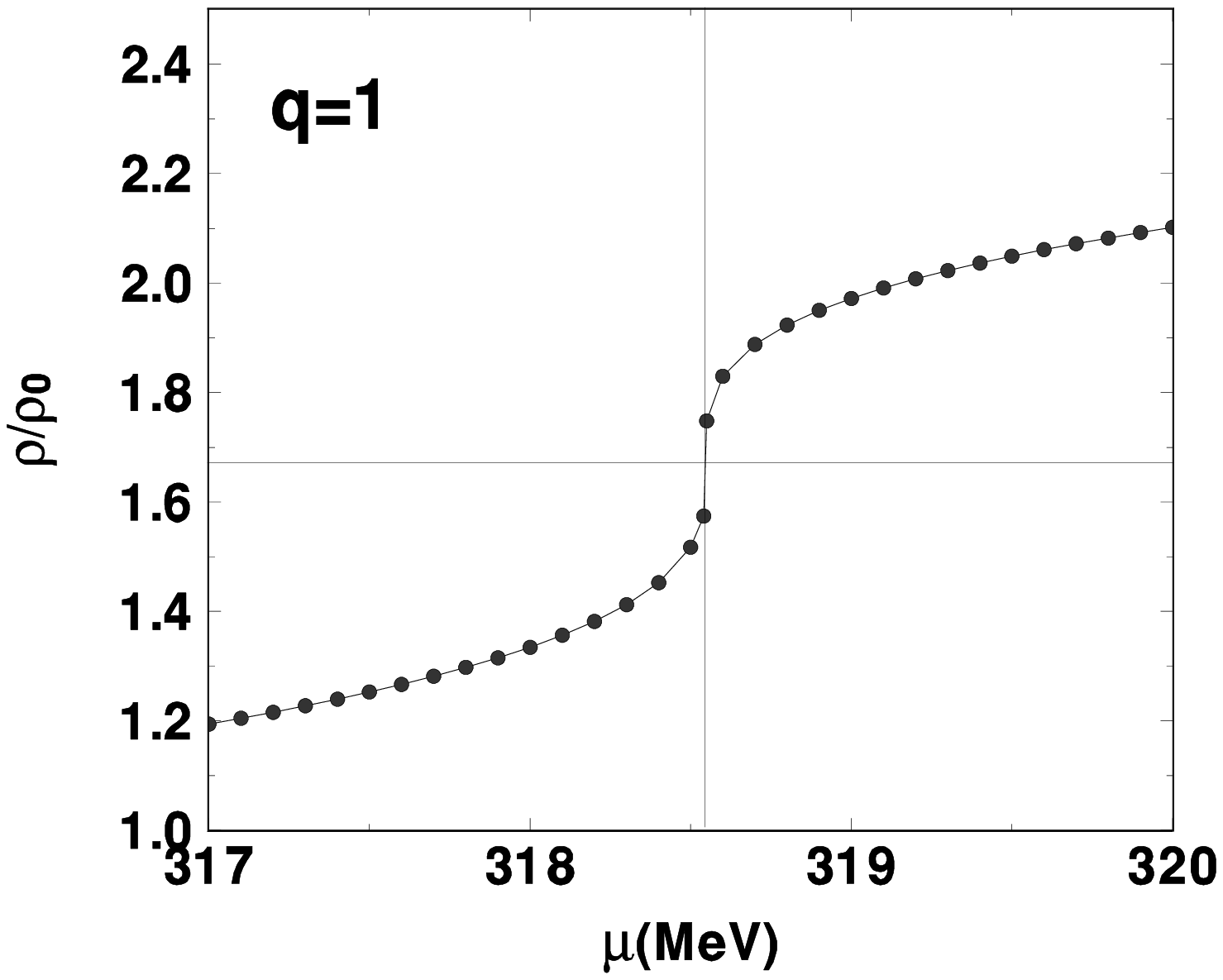}
   \includegraphics[width=5.6cm]{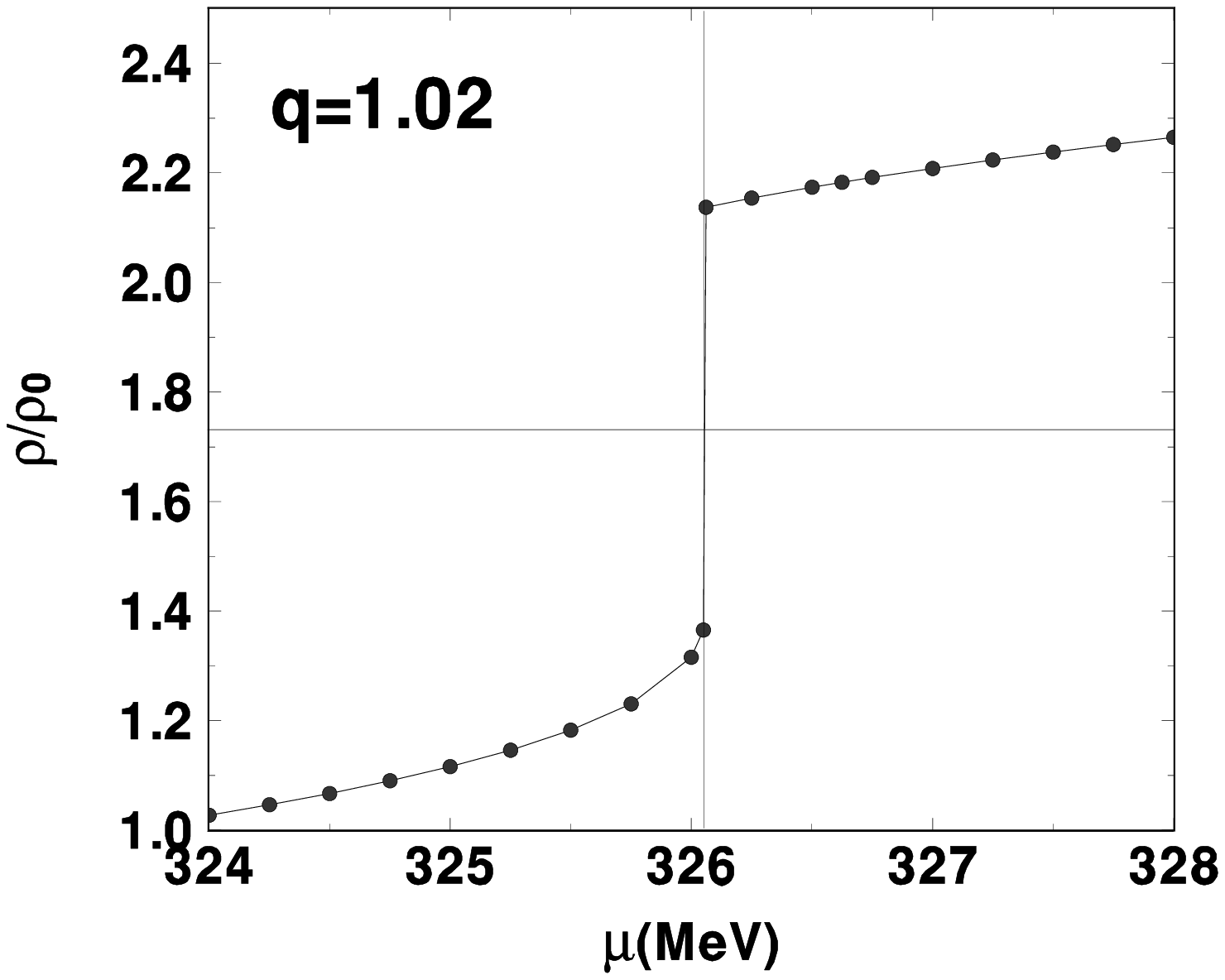}
   \caption{The baryon compression $\rho/\rho_0$ (calculated in the vicinity of the
   critical values of temperature and density indicated by the corresponding
   dotted lines) as function of the chemical potential $\mu$ for different values of the
   nonextensivity parameter, $q =0.98, 1.00, 1.02$. The summary presented in the
   top-left panel is detailed in the three consecutive panels.}
   \label{Figure2}
  \end{center}
\end{figure}
As example of how such approach works we present in Fig.
\ref{Figure1} the pressure at critical temperature $T_{cr}$ as a
function of compression $\rho/\rho_0$ calculated for different
values of the nonextensivity parameter $q$ (see \cite{RW} for more
details). We see that for $q<1$ the critical pressure is smaller
but for $q>1$ it is bigger than the critical pressure for BG
distribution. According to \cite{RW} it is directly connected to
specific correlation for $q < 1$ and fluctuations for $q > 1$. The
role of these factors is shown in more detail in Fig.
\ref{Figure2}. Notice the remarkable difference for the density
derivative at the critical point: from the smooth transition
through the critical point for $q<1$ to a big jump in density for
critical value of chemical potential for $q>1$. It reflects the
infinite values of the baryon number susceptibility, $\chi_B$:
\begin{equation}
\chi_B=\frac{1}{V} \sum_{i=u,d,s}
(\partial\rho_i/\partial\mu_B)_T=-\frac{1}{V} \sum_{i=u,d,s}
\partial^2\Omega/\partial^2\mu_B \big{|}_{_{_T}}. \label{sus}
\end{equation}
The transition between confined and deconfined phases and/or
chiral phase transition \cite{HK} can be seen  by measuring, event
by event, the difference in the magnitude of local fluctuation of
the net baryon number in heavy ion collision \cite{Hatta}. They
are initiated and driven mainly by the quark number fluctuation,
described here by $\chi_B$, and can survive through the freezout
\cite{Hatta,Gavin}. Consequently, our q-NJL model allows to make
the fine tuning for the magnitude of  baryon number fluctuations
(measured, for example, by the charge fluctuations of protons) and
to find the characteristic for this system value of the parameter
$q$. However, it does not allow to differentiate between possible
dynamical mechanisms of baryon fluctuation. We close by noticing
that using $q$ dependent $\chi_B$ leads to $q$-dependent parameter
$\epsilon$ of the critical exponents which describe the behavior
of baryon number susceptibilities near the critical point
\cite{Ikeda}. Whereas in the mean field universality class one has
$\epsilon=\epsilon'=2/3$  our preliminary results using $q$-NJL
model show smaller value of this parameter for $q>1$, ($\epsilon
\sim 0.6$ for q=1.02) and greater for $q<1$ ($\epsilon\sim0.8$ for
q=0.98). It would be interesting to deduce the corresponding
values of $q$ from different  models and from results on lattice
with different universality classes - we plan to present it
elsewhere.

\end{document}